\documentclass[iop,appendixfloats]{emulateapj}
\usepackage{tablefootnote}
\usepackage{appendix}
\usepackage{amsmath}
\usepackage{amssymb}
\usepackage{natbib}
\usepackage{graphicx}

\begin{document}

\title{A Mass-Dependent Yield Origin of Neutron-Capture Element Abundance Distributions in Ultra-Faint Dwarfs}
\shorttitle{Origin of Neutron-Capture Distributions in Ultra-Faint Dwarfs}
\shortauthors{Lee et al.}

\author{Duane M. Lee\altaffilmark{1}, Kathryn V. Johnston\altaffilmark{1}, Jason Tumlinson\altaffilmark{2}, Bodhisattva Sen\altaffilmark{3}, Joshua D. Simon\altaffilmark{4}}
\altaffiltext{1}{Department of Astronomy, Columbia University, New York City, NY 10027}
\altaffiltext{2}{Space Telescope Science Institute, Baltimore, MD 21218}
\altaffiltext{3}{Department of Statistics, Columbia University, New York City, NY 10027}
\altaffiltext{4}{The Observatories of the Carnegie Institution of Washington, Pasadena, CA 91101}

\begin{abstract}
One way to constrain the nature of the high-redshift progenitors of the Milky Way is to look at the low-metallicity stellar populations of the different Galactic components today. For example, high-resolution spectroscopy of very metal poor (VMP) stars demonstrates remarkable agreement between the distribution of [Ti/Fe] in the stellar populations of the Milky Way halo (MW) and ultra-faint dwarf (UFD) galaxies. In contrast, for the neutron capture (nc) abundance ratio distributions [(Sr, Ba)/Fe], the peak of the small UFD sample (6 stars) exhibits a significant under-abundance relative to the VMP stars in the larger MW halo sample ($\sim$ 300 stars). We present a simple scenario that can simultaneously explain these similarities and differences by assuming: (i) that the MW VMP stars were predominately enriched by a prior generation of stars which possessed a higher total mass than the prior generation of stars that enriched the UFD VMP stars; and (ii) a much stronger mass-dependent yield (MDY) for nc-elements than for the (known) MDY for Ti. Simple statistical tests demonstrate that conditions (i) and (ii) are consistent with the observed abundance distributions, albeit without strong constraints on model parameters. A comparison of the broad constraints for these nc-MDY with those derived in the literature seems to rule out Ba production from low-mass SNs and affirms models that primarily generate yields from high-mass SN. Our scenario can be confirmed by a relatively modest (factor of $\sim 3-4$) increase in the number of high-resolution spectra of VMP stars in UFDs.
\end{abstract}


\maketitle

\section{Introduction}\label{intro}
Our understanding of galaxies forming in a hierarchical universe suggests that a large fraction -- and possibly the majority -- of stars now in the halo of the Milky Way (MW) originally formed in smaller separate systems that were subsequently accreted and disrupted by our Galaxy \citep[as originally proposed by][]{searle78}, with the remainder formed {\it in situ} within the main Galactic progenitor \citep{eggen62,abadi03a,abadi03b,zolotov10,mccarthy12a,tissera12}. While the relative contributions of {\it accreted} and {\it in situ} populations remain uncertain, simulations in which the stellar halo is {\it assumed} to be formed entirely by accretion \citep{bullock05,cooper10} have been shown to have levels of substructure in space, velocities and stellar populations that are broadly consistent with observations \citep{bell08,schlaufman09,xue11}. This raises the following question: to what extent can the small systems that survive today (e.g. the satellite galaxies of the MW) be exploited to understand the properties of the small systems that fell in long ago (i.e. the primordial progenitors of the MW halo)?

One approach to this question is to compare and contrast the chemical abundance patterns of the stars in the stellar halo with those in satellite galaxies. For example, at metallicities [Fe/H] $\gtrsim -2$, stars in the low-mass classical dwarf spheroidals generally have lower $\alpha$-element abundances than halo stars \citep[as seen in compilations by][]{venn04,geisler07}. The observed differences can be explained, in general, by the low star formation rates and efficiencies detected in low mass dwarf spheroidals versus the likely progenitors of most halo stars \citep[see, e.g., review by][]{tolstoy09}. Assuming a continuous star formation history, it is true that for all galaxies there exists an epoch for which no appreciable contributions from Type Ia supernovas (which predominantly produce the decline in [$\alpha$/Fe]) are seen. This means that cosmological and astrophysical effects, which can prematurely quench star formation in galaxies such as reionization \citep{hoeft06} and ram pressure stripping \citep{mayer06}, may determine whether low $\alpha$-abundance ratios appear in systems that are accreted early-on. Therefore, these differences can also be explained within the hierarchical picture of structure formation as a result of star formation histories of the surviving satellites being much more extended than those of the progenitors of the bulk of the Halo \citep{robertson05,font06}.  However, this statement pertaining to late-time evolution still begs the question: to what extent are the {\it progenitors} of the stellar halo similar to the {\it progenitors} of the MW's satellite galaxies? This can be addressed by comparing the abundance patterns of stars found in the ``Very Metal Poor'' (VMP) tail of the metallicity distribution (specifically those VMP stars with [Fe/H] $< -2.5$) which, because of their low metallicities, are supposed to have formed early on in the history of the Universe.

\begin{figure}[th]
   \centering
  \includegraphics[width=0.475\textwidth,angle=0]{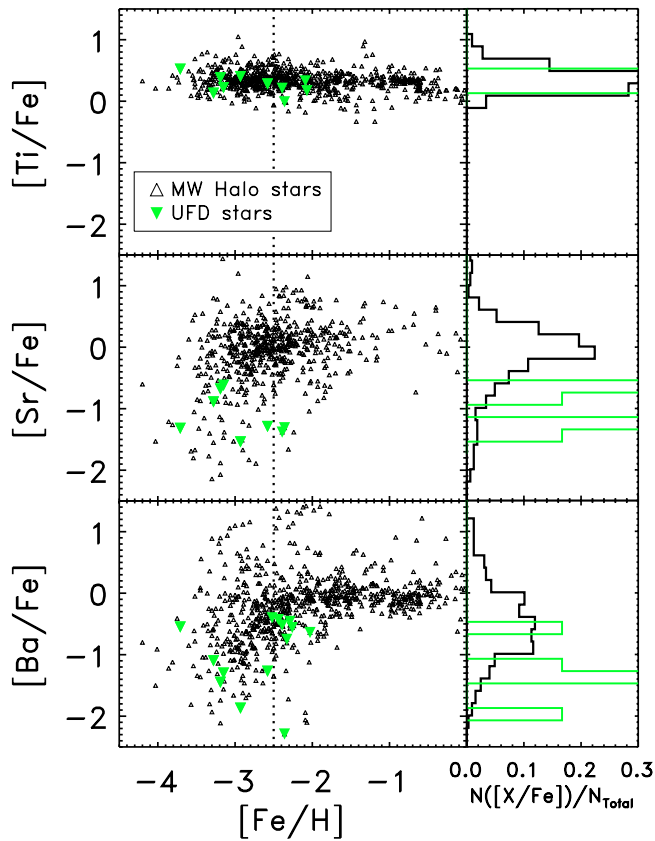} 
   \caption{A compilation of data reproduced from \citep{frebel10} showing a comparison in abundance ratio patterns with Fe for Ti and two nc-elements (Sr, Ba) versus [Fe/H]. Top: Distribution of [Ti/Fe] in MW halo stars (black open triangles) versus UF stars (green upside-down triangles). Middle: Distributions for [Sr/Fe] (nc-element abundance ratio). Bottom: Distributions for [Ba/Fe] (nc-element abundance ratio). Adjacent panels on the right show the relative number distributions, binned by 0.2 dex, for all stars below [Fe/H] = -2.5. Note that stars with upper limits are not included here. See \S \ref{calKS} for details.}
   \label{fig:rproc}
\end{figure}

The open black triangles in the top panel of Figure \ref{fig:rproc} demonstrate that the stellar halo has an average Titanium-to-Iron abundance ratio ([Ti/Fe]) that is roughly constant at all metallicities (measured by [Fe/H]), with a small dispersion that widens in the VMP population. This dispersion can arise when the stochastic nature of star formation is convolved with chemical yields that depend on the masses of the enriching stars \citep{audouze95, ryan96, mcwilliam97, mcwilliam98, norris00}. For example, \cite{karlsson05a} point out that some VMP stars inherit their chemical compositions from gas enriched by just one or a few supernovae (SNe) and have the potential to reflect the full range of abundance ratios implied by the yields from stars of different masses \citep[see also the discussion in][]{karlsson05,tumlinson06a,carigi08, koch09,bland-hawthorn10}. Indeed, the range in [Ti/Fe] exhibited in the stellar halo data at low [Fe/H] is consistent with the predictions for the range in individual yields of Ti from models of exploding stars of different masses \citep{nomoto06,heger10}. In contrast, most stars found with higher metallicities must have been enriched by many SNe, so all their abundances are closer to the average yield for the combined population which can be estimated by integrating the mass-dependent yields (MDY) of the individual stars over the initial mass function (IMF) of enriching stars.

The black open triangles in the lower panels of Figure \ref{fig:rproc} reveal a much wider spread in abundance ratios for the neutron-capture (nc) elements (here, Barium (Ba) and Strontium (Sr)) for VMP stars found in the stellar halo compared to Ti in the upper panel. At these metallicities, \cite{roederer10} suggest (see their Figure 13) that the same massive stars that produce Ti (and the $\alpha$-elements it emulates) also produce nc-elements (thought to originate from core-collapse SN explosions via the {\it r}-process and perhaps from AGB/pre-SN winds via the {\it s}-process) but the forms of the MDYs for Sr and Ba are essentially unknown.\footnote{There may be a few stars that seem to suggest that this relationship breaks down for stellar progenitors with masses $\gtrsim 20 M_{\odot}$. However, \cite{roederer10} state that such stars are probably enriched by unusual SNe at these low metallicites (see their 21$^{st}$ footnote).} 
Hence, one viable explanation of the observed difference in the abundance ratio {\it range} between Ti and nc-elements for VMP stars is to again appeal to the stochastic nature of metal enrichment, but now {\it assume} a much stronger MDY for nc-elements than for Ti.

The green upside-down triangles in Figure \ref{fig:rproc} show abundance ratio measurements in stars in the ultra-faint dwarf satellites (UFD) of the MW \citep[][and references therein]{frebel10b, norris10, simon10, frebel10}. The Ti distributions for the VMP stars in the UFDs (green upside-down triangles) are very similar to the stellar halo (black open triangles), while the nc distributions show a significant difference, with a clear offset between the medians of the two populations that exceeds the spread due to systematic and observational errors \citep{frebel10}.  Several types of ``differences" can be invoked to explain the origin of the galaxy-dependence of these abundance ratio distributions:  
\begin{enumerate}
\item{Differences in the {\bf mixing} of Ti versus nc-elements due to differences in the formation site and process for each element, and, as a consequence, differences in the resultant properties of the enriched ejecta. Assuming that MW progenitors are predominantly larger in size, gas content, and dark matter mass than UFD progenitors, the strength of this effect is mediated by two environmental factors: (i) the depth of the gravitational potential dictates to what extent the different products can be blown out of their respective galaxies by core-collapse SNe; and (ii) the size and dynamics of local gas reservoirs influences how far the products can be evenly mixed in their respective galaxies.} 
\item{Differences in the {\bf IMF} or {\bf MDY} of enriching stars due to preferential enrichment of UFDs from primordial populations of hypernovae \citep{nomoto06} and/or pair-instability SN (PISN) ejecta from Pop III stars \citep[see][and their ``{\it Case B}'' for a discussion of these scenarios]{frebel12}.}
\item{Differences in the {\bf total masses}
of stars enriching the VMP populations in the MW halo and UFD \citep[hereafter the ``stochastic argument,'' similar to {\it Case A} of][]{frebel12}.}  
\end{enumerate}
Note that all of the explanations above implicitly {\it assume} that the UFD progenitors are chemically isolated from MW halo progenitors, which has recently been demonstrated to be a plausible supposition in an analysis of N-body simulations by \cite{corlies13}.

In this paper, we restrict our attention to the last of these ``differences'', which we consider the simplest model possible. 
We extend the discussion of dispersions and skews already in the literature to look at how stochastic chemical enrichment can influence the {\it full} shape of chemical abundance ratio distributions. Our aim is to isolate the influence of this one effect alone. Specifically, we examine to what extent the current abundance ratio distributions of VMP stars in the MW halo and UFDs can be explained {\it without} appealing to differences in mixing, varying the IMF or adopting unique yields. In \S \ref{GA}, we outline and describe the assumptions made in our models. In \S \ref{GE}, we present the general trends in the shapes of abundance ratio distributions produced by our models due to stochastic enrichment. In \S \ref{OC}, we determine the likelihood of drawing the observed distributions of abundance ratios (found in the MW halo and the UFDs) from our simple models. In \S \ref{diss}, we discuss the implications and limitations of our results in connection with expectations from other related studies. Finally, in \S \ref{conc}, we summarize our results and discuss a possible test of the scenario with near-future observations.
\section{General Approach}\label{GA}
Our aim is to determine whether a simple model can simultaneously explain both the {\it similarities} in the distribution of [Ti/Fe] and  the {\it differences} between the distributions of [(Sr,Ba)/Fe] seen for the two systems (the MW halo and the UFDs) represented in Figure~\ref{fig:rproc}. In our model, we assume that: i) the abundance ratios in each observed star represents enrichment from a previous enriching stellar generation (ESG); (ii) the stars within each ESG are sampled from a ``normal" (Salpeter) IMF and produce enrichment with a power-law MDY; and iii) the stellar abundance ratio distributions for each system are the signature of enrichment from an ensemble of ESGs of a characteristic mass, M$_{\rm ESG}$. Note that our simple model assumes that enrichment from Pop III, metal-free stars with peculiar yields does not have a significant effect on abundance ratio patterns at the metallicities observed in UFDs. 

\subsection{Enriching stellar generation}\label{ESG}
Each ESG represents the combined enrichment by stars of total mass $M_{\rm ESG}$ that could be formed in one or many different star clusters. Each ESG realization results from a Monte-Carlo sampling of a \cite{salpeter55} IMF where 
\begin{equation}
\label{eqn:IMF}
\xi=\frac{dN}{dm} = m^{-\alpha} 
\end{equation}
and $\alpha=2.35$. We assume that the lower and upper stellar mass limit for the IMF are $m_{\rm low} = 0.08 M_{\odot}$ and $m_{\rm upp} = 40$, respectively. (In Appendix~\ref{A1}, a range of upper stellar mass limits, $m_{\rm upp} = 30 - 80 M_{\odot}$, are explored.) The lower threshold for stars contributing to chemical enrichment is taken to be $m_{\rm enrich, low} = 8 M_{\odot}$. The number of draws from the IMF is determined by the total and, subsequently, the remaining mass available to form a ESG of $\sim M_{\rm ESG}$. Since this sequence of draws terminates when the total mass drawn exceeds $M_{\rm ESG}$, the actual ESG created only approximates the designated mass. 

\subsection{Stellar enrichment}\label{SE}
Each ESG realization produces a total mass yield for each element $X$ by summation over all individual yields $m_X$ generated from stars of masses $m \ge 8 M_{\odot}$. These yields are determined by a power law of index $\kappa_X$ and normalization $\beta_X$:
\begin{equation}
\label{eqn:mdy}
m_{\rm X} = \beta_{\rm X} \cdot m^{\kappa_{\rm X}}
\end{equation}
In our models, we are assuming that the sources of enrichment are the same in both UFD and MW halo stars. 

Currently, our models only take into account stellar enrichment from massive, short-lived stars which are thought to be the dominant source of enrichment for the VMP populations in both systems. Enrichment by long-lived, low mass stars (excluding binaries) is assumed to become important only at higher metallicities. Although Ba is an archetypical {\it s}-process element at higher metallicities, the trace amounts of Ba observed in the VMP stars we are modeling are produced in core-collapse SNe by the {\it r}-process. There is also a large number of stars with measurable Sr abundances for the same VMP population even though Sr is primarily an {\it s}-process element thought to originate from the AGB phase in low-mass stars (known as the {\it main} {\it s}-process). Therefore, we anticipate that Sr-enrichment in the VMP population comes from a short-lived, but intense, pre-SN/super-AGB phase from massive stars, contributing {\it weak} {\it s}-process elements to the ISM prior to the SN phase \citep{herwig05}; or, perhaps, is simply indicative of {\it r}-process at low metallicities \citep{roederer10}. Recent evidence pointing to fast rotating, massive stars as a viable source for {\it s}-process elements like Sr can be found in \cite{chiappini11}, \cite{frischknecht12}, and references therein. Hence, $m_X$ in our models represents a combined effective yield from both the pre-SN and SN phases of a star of mass m$\ge 8 M_{\odot}$. 

To construct abundance ratios, we first need to account for the common denominator --- Fe abundance. The theoretical yield for Fe tabulated in \cite{nomoto06} varies only slightly over the range of enriching stellar masses examined --- indeed, some previous studies using theoretical Fe yields of $\gtrsim 0.05 M_{\odot}$ have assumed invariant Fe yields for SN ejecta. For consistency with other yields adopted in our models, we set yield parameters for Fe by fitting a power law to the \cite{nomoto06} predictions to find $\beta_{\rm Fe}=$ 0.0607 $M_{\odot}$ and $\kappa_{\rm Fe}= 0.072$.

\begin{figure*}[htbp]
\begin{center}
   \includegraphics[width=0.475\textwidth,angle=0]{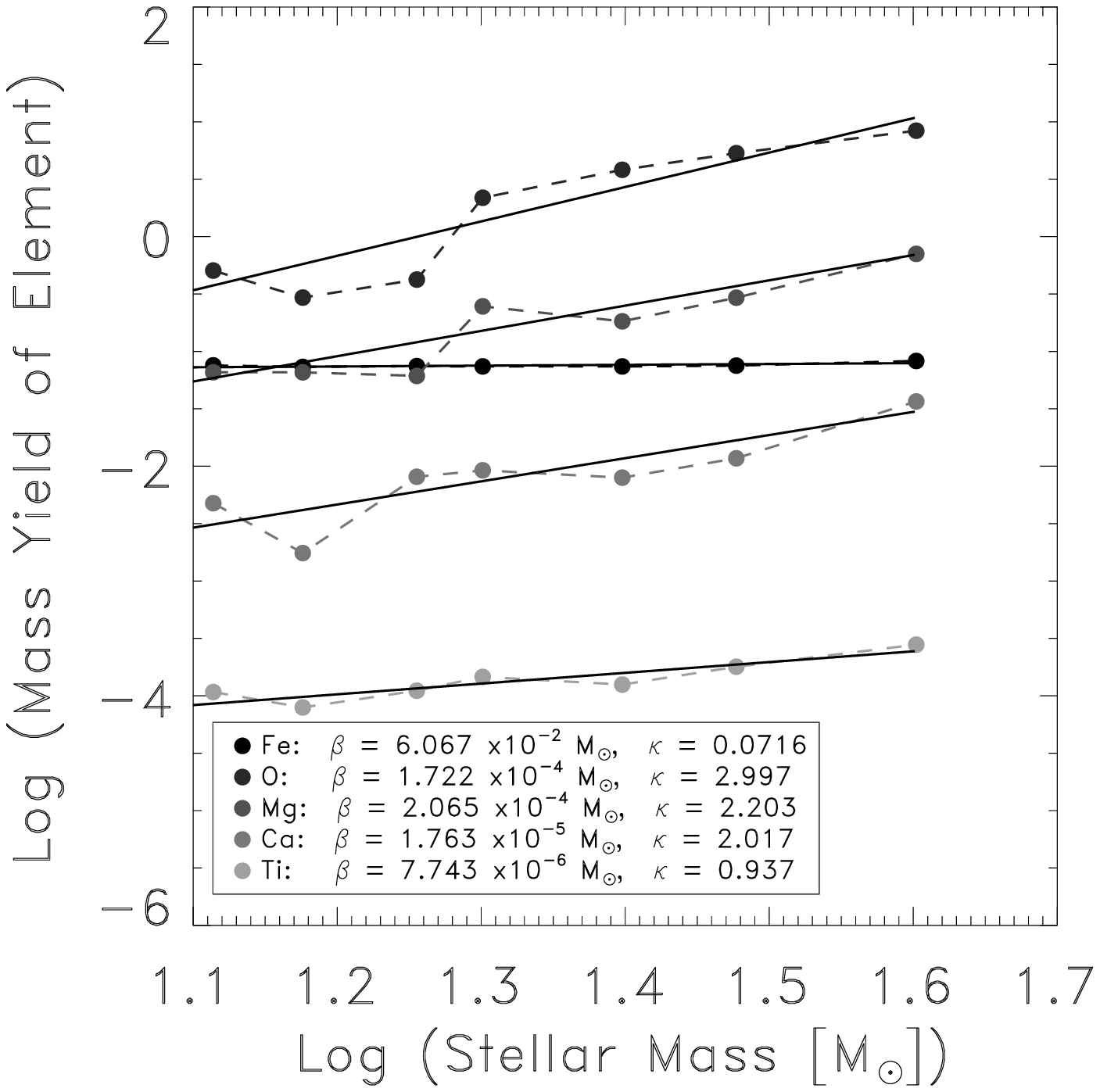}
   \includegraphics[width=0.475\textwidth,angle=0]{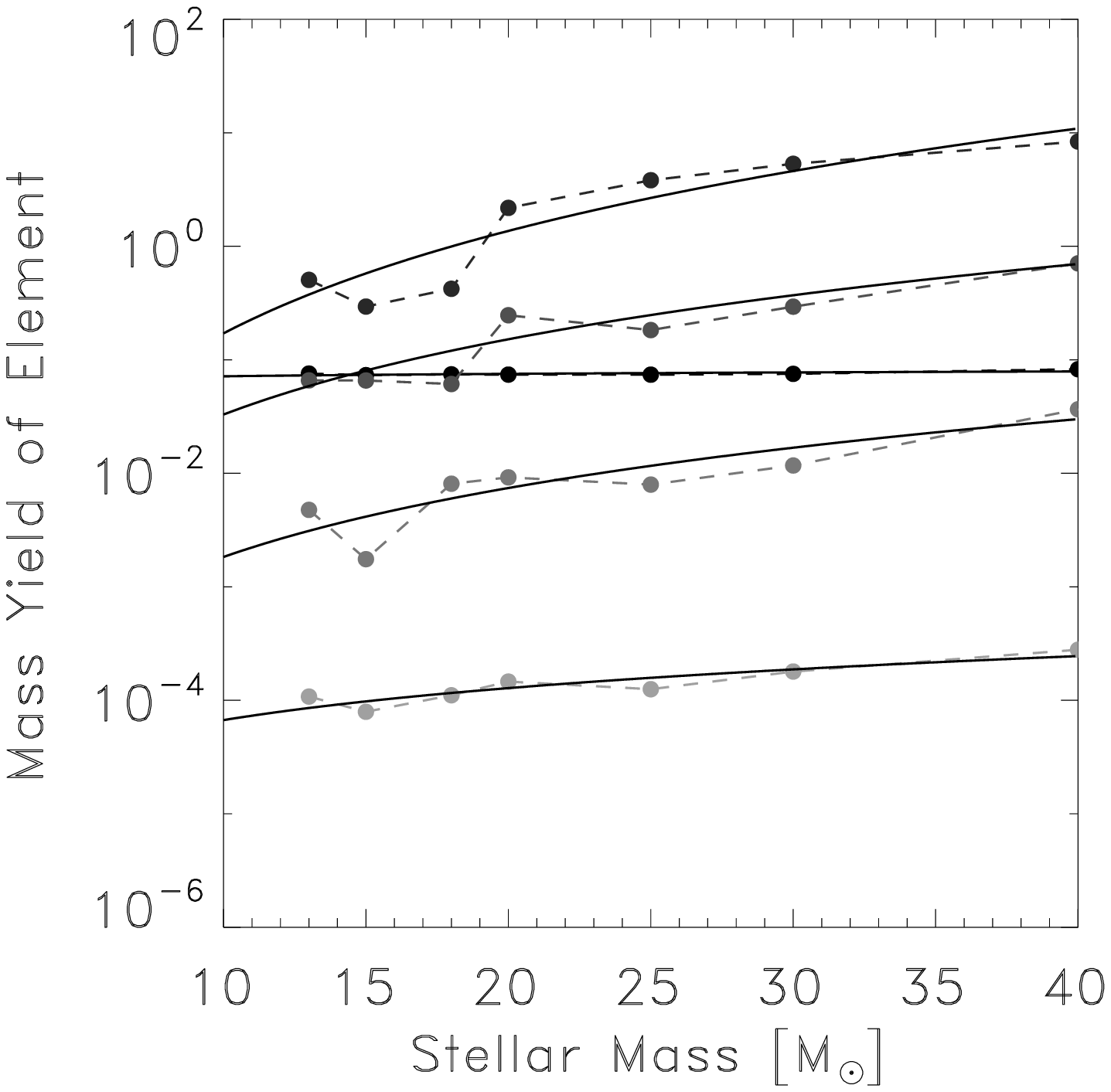}
\caption{Left: Log-log plot of element mass yield versus progenitor stellar mass showing linear fits (black solid lines) with parameters $\beta_{X}$ and $\kappa_{X}$ for some $\alpha$-elements, Ti and Fe MDYs from \citet{nomoto06}. Right: Linear plot of element mass yield versus progenitor stellar mass showing the derived power-law fits (black solid lines) for each element shown in the left plot.}
\label{fig:mdyfits}
\end{center}
\end{figure*}

Figure~\ref{fig:mdyfits} shows our fits to the \citet{nomoto06} theoretical yields at Z = 0.001 ($\simeq Z_{\odot}/18$) for Fe along with fits to archetypical $\alpha$-element MDYs. Also shown is our fit for Ti, which we chose as our {\it known} theoretical MDY because it exhibits the lowest scatter around a power-law fit and has a weak MDY.\footnote{The choice to run models with Z = 0.001 yields versus Z = 0 was arbitrary. However, the difference in MDYs derived for Z = 0.001 ($\kappa_{\rm Fe}= 0.072$ and $\kappa_{Ti}=0.937$) versus Z = 0 ($\kappa_{\rm Fe}= 0.086$ and $\kappa_{Ti}=1.130$) are not significant to this study and use of either set of yields would lead to the same overall results.}

For Ti, the power law fit yields an index of $\kappa_{Ti}=0.937$. The yield normalization $\beta_{Ti}$ is adjusted to maintain agreement between the average abundance ratio calculated for our assumed IMF, {\small $\left<\left[\frac{\rm Ti}{\rm Fe}\right]\right>_{\rm IMF}(\beta_{Ti}, \kappa_{Ti})$}, and average observed abundance ratio, {\small  $\left<\left[\frac{\rm Ti}{\rm Fe}\right]\right>_{\rm OBS}$}, calculated for our VMP ([Fe/H] $<$ -2.5) MW halo sample (see Appendix~\ref{A2}). This adjustment is made to compensate for the failure of the Nomoto models to get the amount of ``fallback'' for Ti correct in their SN explosions (for an explanation, see Figure 12 and \S 8.2 in their paper). 	 

For Sr and Ba, which have no firm yield predictions, we examine a range in $\kappa_{X}$ ($-20\le\kappa_{X}\le 20$) which is wide enough to reveal the relative effects of stochastic sampling in ESGs of different mass (see \S \ref{rTiSr}-\ref{rTiBa}) and allows for a comparison to some proposed yields in the literature (see \S \ref{comdy}). For each $\kappa_{X}$ a $\beta_{X}$ is derived by again requiring a match to the observed average in the MW halo sample, assumed to arise from the fully-sampled IMF.

Finally, it should be noted that while we do track the production of Fe in each ESG realization, we do not explicitly follow evolution in [Fe/H] since the latter is not critical to the scope of this project and would require more detailed assumptions regarding star formation efficiency, mixing, infall, and blowout. 

\subsection{Parent distributions and synthetic ``Child'' samples}\label{parent}
Following the prescription given in \S\ref{SE}, each ESG realization from \S\ref{ESG} produces a chemical abundance ratio for [X/Fe] (where $X$ represents Ti, Ba or Sr), which is supposed to represent a possible enrichment pattern for a subset of the total population of stars that exist in the observed systems. Thus, each ESG produces one enrichment pattern from which many stars can sample. However, the numbers are proportional to how common that enrichment pattern is (as determined by the distribution of patterns from the ESGs generated). For a given set of parameters ($M_{\rm ESG}$, $ \kappa_{\rm X}$) we construct two-dimensional ``parent distributions'' in the [Sr/Fe]-[Ti/Fe] and [Ba/Fe]-[Ti/Fe] planes from ensembles of enrichment by 1000 ESGs. Each parent represents a model for the intrinsic stellar distribution from which we can draw random synthetic samples (``{\it children}'') to compare to the MW halo and UFD observed data distributions. Each {\it child} contains the same number of synthetic stars as the number of observed stars and their stellar abundance ratios are scattered by observational errors which are taken to be 0.15 dex (as a conservative lower bound). 

\section{Results  I: General Effects}\label{GE}
In this section we develop some intuition by examining the effect of varying parameters ($M_{\rm ESG}$, $ \kappa_{\rm X}$) on the shape of the abundance ratio distribution in [$X$/Fe] in one dimension. 
\subsection{Phenomenological expectations}\label{PP}
\begin{figure}[htbp]
\begin{center}
   \includegraphics[width=0.475\textwidth,angle=0]{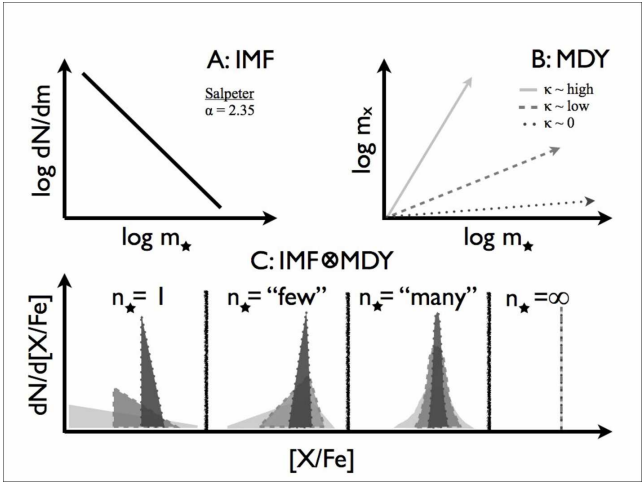}
\caption{A schematic displaying the assumptions of our model and the various effects that arise from convolving the IMF, MDY, and the number of enriching stars per ESG, n$_{\star}$, together. Note that n$_{\star}$ is not exactly proportional to the characteristic ESG mass, $M_{\rm ESG}$, due to stochastic sampling of the IMF. However, n$_{\star}$ does give some indication of the magnitude of $M_{\rm ESG}$. Panel {\bf A} shows a plot of the Salpeter IMF used in our models, indicating a large ratio of low-to-high mass stars produced in ESGs. Panel {\bf B} displays three different MDY ``strengths'' measured by the slope $\kappa_{\rm X}$ as indicated by approximately zero (dotted), low (dashed), and high (solid) labelled lines. These MDY strengths result in the trends we expect to find in abundance ratio distributions for VMP stars (shown in panel {\bf C}). Panel {\bf C} shows the three types of distributions that can arise for different positive MDYs  resulting from the convolution of the IMF and MDYs for four different characteristic values of n$_{\star}$. The shades/outlines of the distributions represent the ``strength'' of the MDY (as shown in panel {\bf B}): $\kappa\sim0$ (dark grey; dotted line), $\kappa\sim$ {\it low} (grey; dashed line), and $\kappa\sim$ {\it high} (light grey; solid line).}
\label{fig:scheme}
\end{center}
\end{figure}

Figure~\ref{fig:scheme} illustrates schematically the trends we expect to see in our distributions resulting from the combination of the IMF, the MDY($\kappa_{\rm X}$), and the number of enriching stars, n$_{\star}$, generated in a ESG (which is proportional, on average, to $M_{\rm ESG}$). 

In panel A, the Salpeter IMF is shown, illustrating that many more lower mass stars are produced for a given number of high mass stars in any ESG. This property is generic to all proposed IMFs in nearby galactic environments investigated in the literature \citep{kroupa02,chabrier03,elmegreen05,elmegreen06a,elmegreen07}.

In panel B, the MDY for various $\kappa_{\rm X}$ are shown: an approximately constant mass yield across all stellar masses ($\kappa_{\rm X} \simeq 0$), a small/weak change in mass yield  (low $\kappa_{\rm X}$ values), and a large/strong change in mass yield (high $\kappa_{\rm X}$ values). It should be noted that these power law fits are a rough 1$^{st}$-order approximation to the non-monotonic functions for MDYs anticipated in nucleosynthetic yield models \citep[e.g.][]{nomoto06,heger10} for both Ti and nc-elements. The detailed shape of these functions will be another key factor which contributes to the range and shape of observed abundance ratios, but is not considered in this paper to keep our models as simple as possible (and because the mass-dependence of stellar yields for most elements is not well understood at present).

In panel C, trends in the distribution of yields from an ensemble of enriching ESGs as a result of combining the IMF with MDY (IMF$\otimes$MDY) are shown for different numbers of enriching stars per ESG, n$_{\star}$. 

In the limit of n$_{\star}$= $\infty$ (right hand plot of panel C) complete sampling of the IMF is achieved, resulting in a single mean value 
\begin{equation}
\label{eqn:xfe}
\left<\left[\frac{\rm X}{\rm Fe}\right]\right>_{\rm IMF} = \left<\left[\frac{\rm X}{\rm Fe}\right]\right>_{\rm OBS}
\end{equation}
for all realizations.

In the opposite limit of n$_{\star}$=1 (left-hand plot of panel C), we expect to directly sample the full range of yields contributed from individual stars, with frequencies dictated by the IMF. Hence a strong MDY (high $\kappa_{\rm X}$; solid line/light-shaded area) will produce a wide distribution while a weak MDY (low $\kappa_{\rm X}$; dotted line/dark-shaded area) will produce a narrow one. For positive $\kappa_{\rm X}$, the skew of these distributions will be positive or right-skewed, meaning that their extended tails are found to the right of the median and peaks are found to the left. In the case of negative $\kappa_{\rm X}$ (not shown), the skew of the distributions will become negative, with the extended tail to the left of the median. A wide range of distributions can be observed between these two limits. For an element X with large, positive $\kappa_{\rm X}$ (solid lines and light-gray areas in Figure~\ref{fig:scheme}), various distributions can be exhibited depending on the value of n$_{\star}$. 

For example, with n$_{\star}$=``few'',  the convolution of yields with the IMF from a few enrichers can generate negatively-skewed (left-skewed) distributions.\footnote{This tendency is modulated by the specific number of stars, the strength of the MDY, and the upper-limit of the IMF within this range. Therefore, positively-skewed and gaussian-like distributions are not necessarily excluded.} 
Although massive enrichers are found less frequently than their lower mass counterparts, their individual chemical yields can dwarf those contributed by lower mass stars.  Hence, the orientation of the tail of the distribution can flip compared to the $n_\star=1$ case due to the weighted contribution of the ``few'' high mass enrichers with large absolute yields.

For n$_{\star}$=``many'', the average number of n$_{\star}$ realized in each ESG is high enough to start altering the distribution from a poisson-like distribution to a gaussian-like distribution via the {\it law of large numbers}. This effect arises from a counter-balance between the plentiful, although low impact, low-mass enrichers and the sparse, yet high impact, high-mass enrichers which leads to an ``erosion'' of possible abundance ratios at the margins of the distribution (homogenization), thus narrowing the distribution in accordance with the {\it central limit theorem}. 

\subsection{Model distributions}\label{MD}
We can assess the validity of our phenomenological expectations, given in \S \ref{PP}, by examining ensembles of many ESGs realized with identical parameters, to create chemical abundance ratio probability distributions. The features of interest are systematic changes in the: 1) {\it variance} (dispersion), 2)  {\it skewness} (lopsidedness), and 3)  {\it kurtosis} (peakedness) of the distribution. As noted in \S \ref{SE}, the ``means'' of our distributions are set by the observed average abundance ratio but these higher moments emerge from the parameters specified for $\kappa_{\rm X}$ and $M_{\rm ESG}$. 
\begin{figure*}[htp]
   \centering
   \includegraphics[width=0.95\textwidth,angle=0]{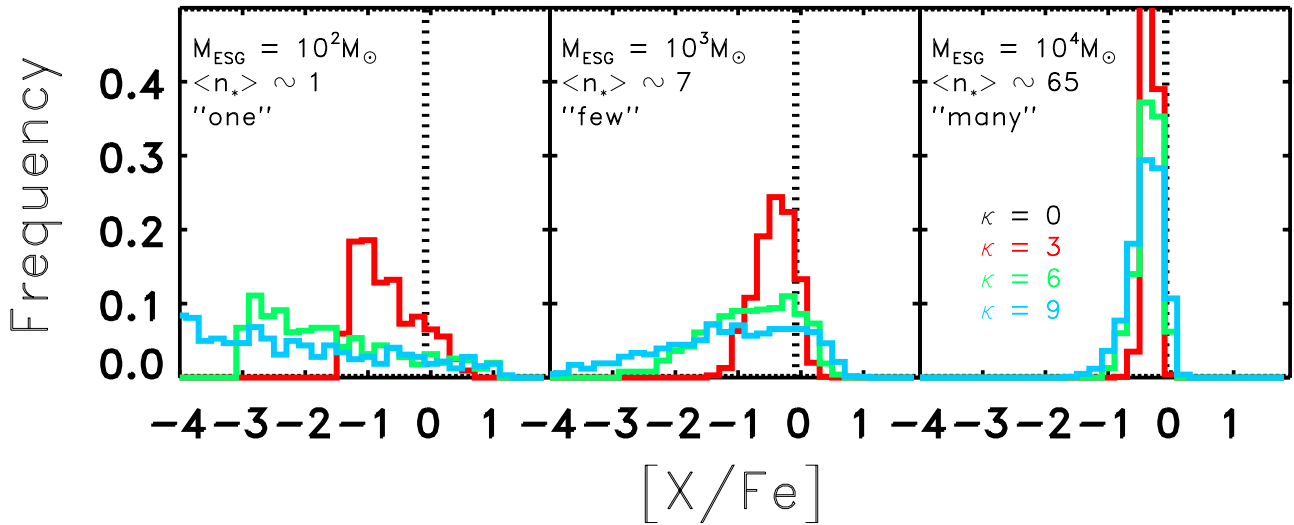}
   \caption{Distributions of abundance ratios produced from 1000 realizations of an ESG, with $M_{\rm ESG} = 10^{2} M_{\odot}$ (left panel), $10^{3} M_{\odot}$ (middle panel), and $10^{4} M_{\odot}$ (right panel). Color of the distribution refers to the corresponding $\kappa_{X}$ used for the MDY: 3 (red), 6 (green) and 9 (blue). The black vertical dotted line shows the average for all ESGs with $\kappa_{X}$ = 0. The average number of enriching core-collapse supernovas are represented by $<$$n_{\star}$$>$.}
   \label{fig:models}
\end{figure*}

\begin{figure*}[tp]
   \centering
   \includegraphics[width=0.95\textwidth,angle=0]{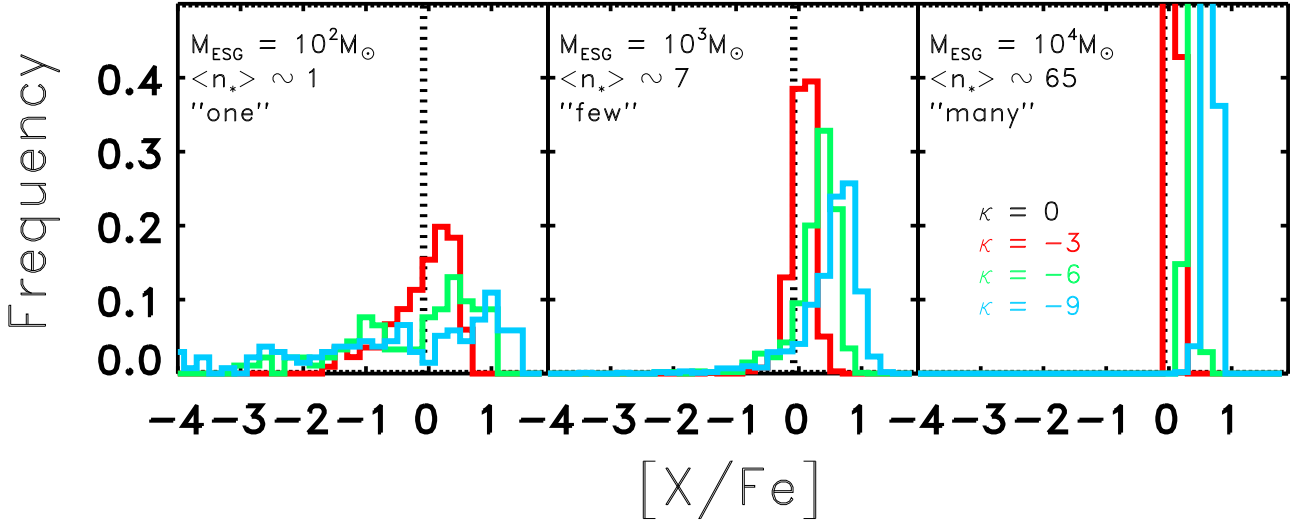}
   \caption{Figure is the similar to Figure~\ref{fig:models} but shows distributions derived from negative MDYs.}
   \label{fig:negmodels}
\end{figure*}

Figures~\ref{fig:models} and~\ref{fig:negmodels} illustrate the general trends found for various parameters ($\kappa_{\rm X}$, $M_{\rm ESG}$). Figure~\ref{fig:models} shows a number of features in these distributions that are similar to both our schematic framework and the observed distributions. Each panel corresponds to a different decade in $M_{\rm ESG}$ ($= 10^2, 10^3, 10^4 M_\odot$ respectively) realized 1000 times to create distributions with average number of enriching stars given by $<$$n_{\star}$$>$ $\simeq1$, $\simeq7$, and $\simeq65$, analogous to the one, ``few'', and ``many''  enrichers in the schematic in Figure \ref{fig:scheme}. Comparing the different colored histograms within each panel, increasing the value of $\kappa_{X}$ = 3 (red), 6 (green) and 9 (blue) leads to a broadening of the distribution. The black vertical dotted line shows the average for the distributions which correspond to yields from all ESGs generated with $\kappa_{X}$ = 0. Comparing the same colored histograms across the panels, shows stochastic effects producing right-skewed distributions for  $n_\star\sim 1$ (left panel), left-skewed distributions for $n_\star \sim$ ``few'' (middle panel) and gaussian-like distributions for $n_\star \sim$ ``many'' (right panel). 

Figure~\ref{fig:negmodels} shows the general trends for negative MDYs. A comparison between Figure~\ref{fig:models} and Figure~\ref{fig:negmodels} demonstrates that distributions derived with positive MDYs for X cause negatively or positively skewed distributions while negative MDYs only lead to negatively skewed distributions.

Most importantly, if we compare Figure~\ref{fig:models} to the observations (``by eye'') we see that the simple stochastic picture can be supported if the following criteria are met:
\begin{itemize}
\item Abundance ratios in MW halo stars reflect enrichment by ESG with masses sufficient to produce ``few''-to-``many'' enrichers, while abundance ratios in UFD stars reflect enrichment by ESG with masses that would produce roughly ``one'' enricher; 
\item Ti is well approximated by low $\kappa_{\rm X}$, resulting in a similar distribution for any $n_{\star}$ (or $M_{\rm ESG}$);
\item nc-element yields are well approximated by high $|\kappa_{\rm X}|$, resulting in noticeably different distributions for low versus high $n_{\star}$ (or $M_{\rm ESG}$).  
\end{itemize}

\section{Results II: Observational Constraints On Model Parameters}\label{OC} 
Having demonstrated in principle that skewed abundance ratio distributions can be obtained when incomplete sampling of the IMF is coupled with strong MDYs we now assess whether this explanation is sufficient to explain the current observed samples.

\subsection{Selecting a comparison sample}\label{ONC}
As stated earlier, our models were designed to track stellar abundance ratios that originate from the evolution of {\it high-mass} stars that are not in binary systems (i.e., the combined yields from a super-AGB/pre-SN phase and/or post-SN wind). 
Hence we select our sample from the \cite{frebel10} compilation to exclude stars whose abundance ratios are likely to include enrichment from other sources. 

Specifically, abundance ratio contributions from {\it low-mass} stars (e.g. AGB winds or Type Ia SNe) are limited by looking at VMP stars (with [Fe/H] $<$ -2.5 in our case): because of their low-metallicity, VMP stars are assumed to have formed before long-lived low-mass stars had a chance to contribute significantly to chemical enrichment \cite{vargas13}. 

\begin{figure}[bhp]
    \centering
    \includegraphics[width=0.475\textwidth,angle=0]{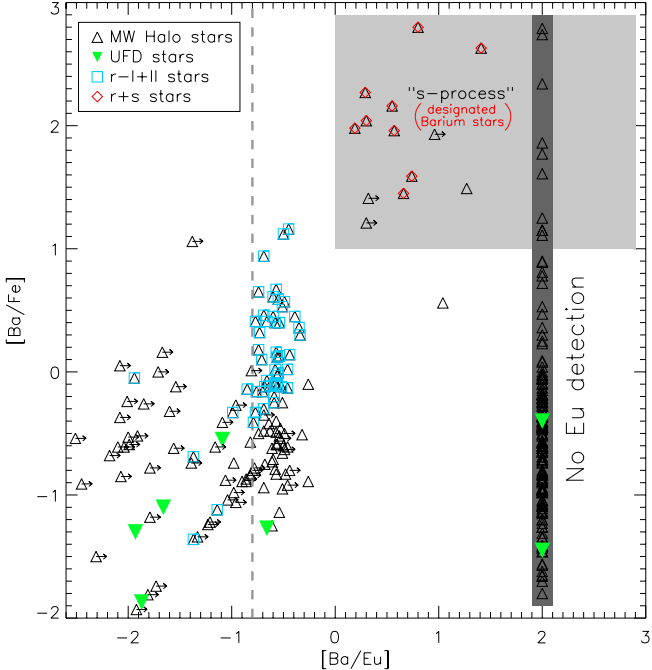} 
    \caption{[Ba/Fe] vs. [Ba/Eu] for the VMP (below [Fe/H] = -2.5) stars in the  MW halo/UFD data set. Black triangles are MW halo stars without given {\it r}- or {\it s}-process abundance ratio designations. Blue squares refer to stars with {\it r}-process abundance ratios (either class I or II) and red diamonds refer to {\it r}+{\it s}-process stars as designated by \cite{frebel10}. The pure {\it r}-process upper-limit, designated Barium stars (exclusion) region, and stars with non-detections of Europium and identified by a dashed grey line, grey rectangular region, and dark grey stripe, respectively (see text for explanation).}
    \label{fig:dia}
 \end{figure}

In addition, we use Figure~\ref{fig:dia} to exclude stars whose abundance patterns could reflect enrichment during binary evolution by identifying those stars that fall within the abundance ratio boundaries of [Ba/Fe] $> 1.0$ dex and [Ba/Eu] $> 0.5$ dex \citep[indicated by the grey rectangular region, from the diagnostic prescription listed in the review by][]{beers05}. 
These ``Barium stars'' are thought to be produced during binary evolution from {\it s}-process-enhanced Barium enrichment in the common envelope \citep{smith90,mcclure90} or wind accretion \citep{boffin88} phases.
The validity of this simple diagnostic is confirmed by the locations of the stars highlighted in red and blue which indicate where \citet{frebel10}, using a more detailed abundance ratio analysis, designated stars as enriched by both {\it r}+{\it s}-process (in red) or by {\it r}-process (either class I or II) alone (in blue). For those stars with no Eu detection (vertical grey stripe) we also exclude those stars with [Ba/Fe] $>$ 1.0 dex since a non-detection for Eu ensures that [Ba/Eu] $>> 0.5$ dex.

\begin{figure}[tbh] 
    \centering
    \includegraphics[width=0.475\textwidth,angle=0]{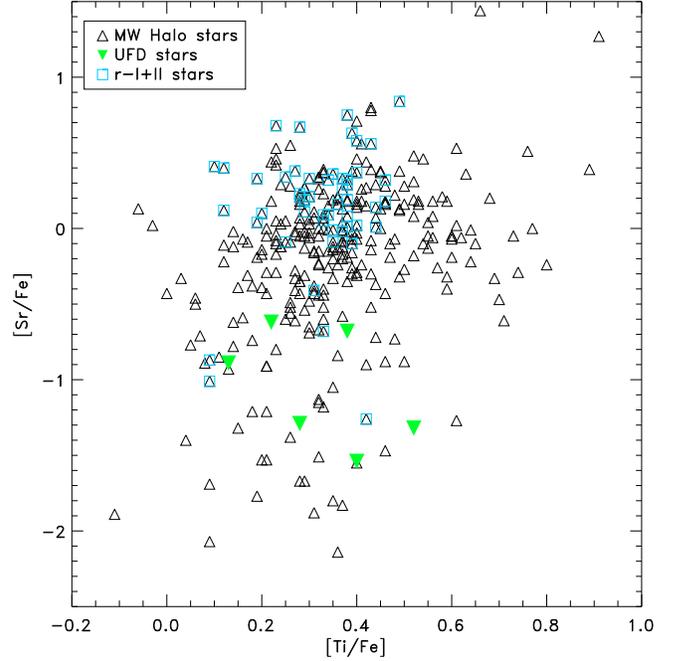} 
    \caption{[Sr/Fe] vs. [Ti/Fe] for our compiled observed MW halo/UFD data set. Symbols are the same as those defined in Figure~\ref{fig:dia}.}
    \label{fig:srti}
 \end{figure}

\begin{figure}[tbh] 
    \centering
    \includegraphics[width=0.475\textwidth,angle=0]{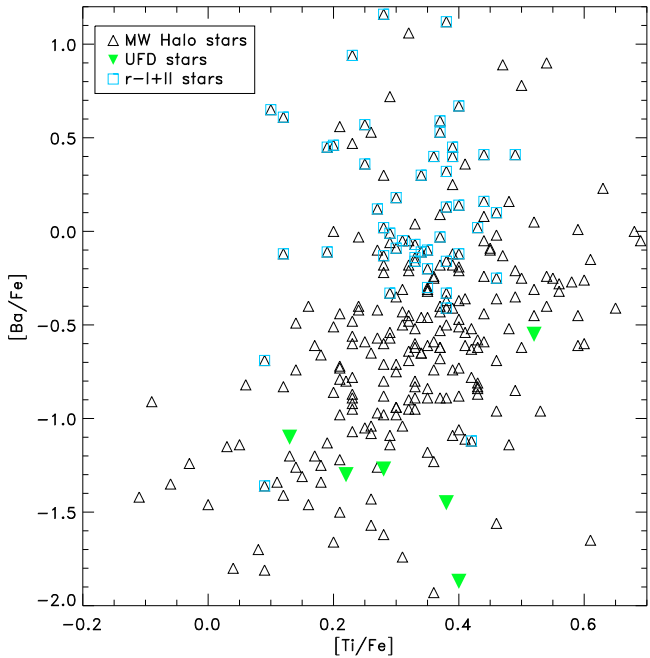} 
    \caption{[Ba/Fe] vs. [Ti/Fe] for our compiled observed MW halo/UFD data set. Symbols are the same as those defined in Figure~\ref{fig:dia}.}
    \label{fig:bati}
 \end{figure}

Figures~\ref{fig:srti} and~\ref{fig:bati} display our final samples in the [Sr/Fe]-[Ti/Fe] and [Ba/Fe]-[Ti/Fe] planes containing 322 stars ($n_{\rm MW}=316$ and $n_{\rm UF}=6$) and 269 stars ($n_{\rm MW}=263$ and $n_{\rm UF}=6$), respectively. As noted above, these samples are limited to stars not designated as ``Barium stars'' with [Fe/H] $< -2.5$. In addition, only stars with values for both elements (either Sr and Ti or Ba and Ti) that are definitively measured are included (i.e. excluding upper limits).

\subsection{Comparing data and models with a ``paternal likelihood test''}\label{calKS}
To directly compare our models to observations, we construct a test to determine the likelihood that the observed stellar abundance ratio samples for the MW halo or UFDs, shown in Figures~\ref{fig:srti} and~\ref{fig:bati}, could be drawn from the 2-d parent distributions generated by a particular parameter set (see \S \ref{parent}). Our {``}paternal-likelihood test,{''} is built around the comparison of our samples to each {\it parent} using the {\it D}-statistic derived from the two-dimensional Kolmogorov-Smirnov (2dKS) test \citep{press92}. The {\it D}-statistic represents the maximum difference (supremum) between two cumulative distribution functions (CDFs) --- a smaller supremum indicates a higher likelihood that both CDFs are drawn from the same population. 

While the values of the {\it D}-statistic can be used to rank our parameter sets given the observed data, the 2dKS test alone is insufficient for our purposes. The multitude of possible data orderings used to create CDFs in multi-dimensional samples \citep{peacock83,fasano87} means that the {\it D}-statistic cannot be simply converted to a likelihood in a model-independent manner. This problem is particularly challenging given the small number of stars (6) used in the UFD samples where large differences in {\it D}-statistics between parameter sets may not actually represent significantly different likelihoods. Our {\it paternal-likelihood test} addresses this limitation by generating child-parent distances (${\rm {\it D}_{\rm cp}}$) for a large number of synthetic {\it child} samples (with sample sizes equaling the observed data size) drawn (bootstrapped) from the parent. The distribution of ${\rm {\it D}_{\rm cp}}$ can then be used to assess the likelihood of observing the distance ${\rm {\it D}_{\rm dp}}$ between the collected data samples and the parent.

\begin{figure}[tbp]
   \centering
   \includegraphics[width=0.475\textwidth,angle=0]{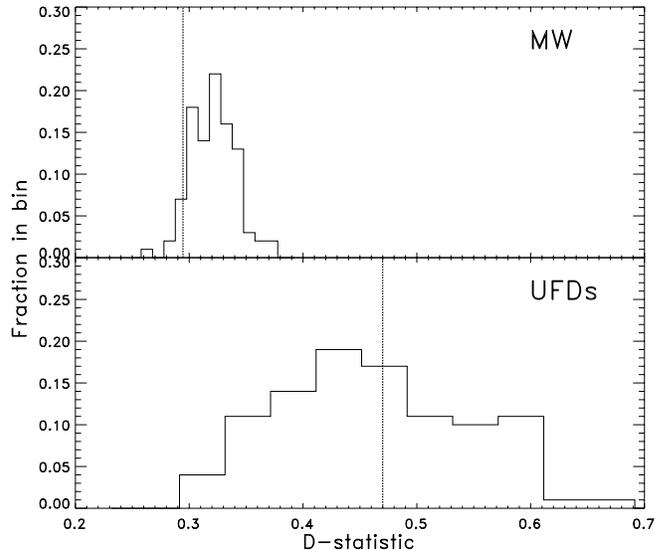}
   \caption{Histograms show the distributions of the child-parent D-statistic, $D_{\rm cp}$, for ``children'' with the same size as the observed Ba data sets ($n=316$ for MW, upper panel; $n=6$ for UFDs, lower panel) drawn from parents with model parameters $\kappa_{X}$ = 9.5, ${M_{ESG}}^{MW}$ = $10^{3.5} M_\odot$ (top) and ${M_{ESG}}^{UF}$ = $10^{2.0} M_\odot$ (bottom). Bin sizes equal $|${\it D}$_{Max}$-{\it D}$_{Min}$$|/10$ in {\it D} range. The vertical line marks the D-statistic for the observed data sets, ${\rm {\it D}}_{\rm dp}$.}
   \label{fig:results0}
\end{figure}

Specifically, we generate $n_{\rm children}= 100$ from each {\it parent} (defined by parameters $M_{ESG}$, $\kappa_{\rm X}$, $m_{\rm upp}$). Each {\it child} is comprised of {\it n} randomly-sampled stellar abundance ratios from the parent distribution where {\it n} equals the number of observed stars from the observed comparison data sample. Figure~\ref{fig:results0}, for example, shows a distribution of {\it D}-statistic ranks calculated for the [Ti/Fe]-[Ba/Fe]-plane using children drawn from one of our parent distributions to assess parental likelihood for the MW halo (upper panel) and UFDs (lower panel), respectively.  The spreads in the distributions are influenced by both the observational/systematic errors and the sample size. As expected, a larger sample of stellar abundance ratios increases our certainty about the likely parent of the observed distribution. 

We assess the significance of the comparison rankings between the observational data and the {\it parent}, $D_{\rm dp}$ (indicated by vertical dashed line in Figure~\ref{fig:results0}) by calculating a {\it p}-value --- i.e. the fraction of {\it children} that are ranked as more different from the parent than the observed data (shown as the fraction of the histogram that lies to the right of the vertical line in Figure~\ref{fig:results0}):
\begin{equation}
\label{pval}
{\rm {\it p}-value} = \frac{n_{\rm children}({\rm {\it D}}_{\rm cp} > {\rm {\it D}}_{\rm dp})}{n_{\rm children}}.
\end{equation}
The higher the {\it p}-value, the more likely the observed abundance ratios are a potential ``offspring'' of the parent. 

\subsection{Results from the [Ti/Fe]-[Sr/Fe] plane}\label{rTiSr} 
\begin{figure}[t]
   \centering
   \includegraphics[width=0.475\textwidth,angle=0]{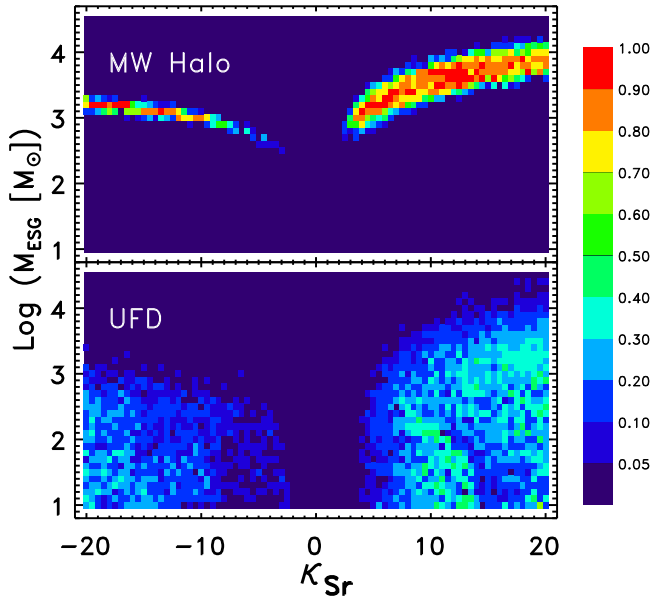}
   \caption{Figure shows the likelihood ({\it p}-value) distribution for the MW halo (upper panel) and UFDs (lower panel) derived from different models with $M_{\rm upp}= 40 M_{\odot}$ (reflecting the [Ti/Fe]-[Sr/Fe] plane) as a function of $M_{\rm ESG}$ and $\kappa_{\rm X}$ for Sr yields. See text for explanation of features.}
   \label{fig:results1a}
\end{figure}

Figure~\ref{fig:results1a} summarizes the results of our paternal likelihood test applied to the MW halo (upper panel) and UFD (lower panel) samples in the [Ti/Fe]-[Sr/Fe] plane. The color of the plot indicates the likelihood (i.e. the {\it p}-value) of the observations being drawn from a parent of particular $M_{\rm ESG}$ and $\kappa_{\rm X}$, and for a fixed $M_{\rm upp}=40 M_{\odot}$\footnote{We find that some spurious likelihoods can arise from models that have sample dispersions of $\sim$0.3 dex or less (i.e., on par with the observational or systematic errors). These artifacts are caused by a limitation in the way the 2dKS test handles models with a relative dearth of data sampled in the wings of its distribution (see \cite{babu06,babu04,stephens74} for an explanation). Models with intrinsic dispersions of $\simeq 0$ are emblematic of this limitation.  Fortunately, such models can be trivially identified (by their aforementioned dispersions) to be incompatible with the observed data and are therefore recorded with likelihoods of less than 5\%.}.

From the upper panel it is immediately apparent that models with $M_{\rm ESG} \gtrsim 10^{3} M_{\odot}$ are preferred in generating MW halo-like distributions. Furthermore, these models are consistent with a wide range of $|\kappa_{\rm X}| \gtrsim 2$ values due to the degeneracy between stochastic sampling of the IMF (governed by $M_{\rm ESG}$) and the effect of varying the strength of the MDY: the IMF is more completely sampled as $M_{\rm ESG}$ gets larger which will tend to homogenize the stellar abundance ratios, but this effect can be compensated for with a higher MDY strength in order to maintain a sufficient width to match the MW halo distribution. 

Differences between the location and width of the trends apparent in the upper panel for $\pm\kappa_{\rm Sr}$ can be attributed to the relative weighting of low/high-mass enrichers in each case. Since there are significantly more low-mass enrichers than high-mass enrichers generated for $M_{\rm ESG} \gtrsim$ a few hundred solar masses, homogenization is reached sooner for negative $\kappa_{\rm X}$ (i.e. at a lower ESG mass) than for ESGs with a positive $\kappa_{\rm X}$. Also, the smaller width of the probability distribution for $\kappa_{\rm X} < 0$ reflects the diminished contributions of high mass stars because they are (in this scenario) both rare and have yields that are small relative to their less massive counterparts, thus shrinking the range of $M_{\rm ESG}$ capable of producing the observed MW halo distribution.

The lower panel displays the results of the same analysis for the six stars in the UFD sample. The two regions of significant likelihood are analogs to the negative and positive $\kappa_{\rm X}$ trends found for the MW halo, but the paucity of observed stars in the UFD sample means that a much broader set of models are compatible with the observed chemical distributions. Therefore, we see models with substantial likelihoods ({\it p}-values) across more than two decades in $M_{\rm ESG}$ for a variety of $\kappa_{\rm X}$ values. Despite the breadth of possible solutions found in each panel, they demonstrate (as a whole) that our simple model of stochastic enrichment is sufficient to explain the Ti and Sr abundance ratio distributions in the MW halo and UFDs simultaneously, provided that: (i) the UFD systems were enriched by a lower ESG mass than the progenitors to the MW halo stars; and (ii) Sr yields can be characterized by a power law with a relatively larger $|\kappa_{\rm X}|$ when compared to Ti yields. 

\subsection{Results from the [Ti/Fe]-[Ba/Fe] plane}\label{rTiBa}
\begin{figure}[t]
   \centering
   \includegraphics[width=0.475\textwidth,angle=0]{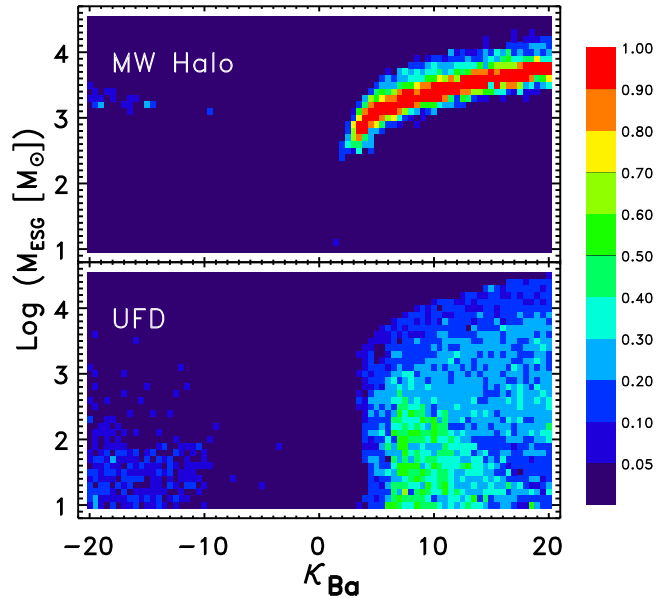}
   \caption{Figure is the same as Figure~\ref{fig:results1a} for Ba yields.}
   \label{fig:results1b}
\end{figure}

Figure~\ref{fig:results1b} summarizes our analysis of model comparisons to samples observed in the [Ti/Fe]-[Ba/Fe] plane. This figure offers additional confirmation of the results from the [Ti/Fe]-[Sr/Fe] plane: that the same simple model of stochastic enrichment with the same masses for MW halo and UFD enrichers preferred can also explain the distributions in this plane. The UFD results here suggest a slightly lower $\kappa_{\rm X}$ for the MDY of Ba compared to Sr. Also, in the case of Ba, a negative MDY seems highly unlikely from our analysis. This result can be explained by comparing the UFD distributions from Figures~\ref{fig:srti} and~\ref{fig:bati} to the $M_{ESG}=10^{2} M_{\odot}$ models from Figure~\ref{fig:negmodels}. It is apparent that a smaller negative offset along with a high concentration of abundances is favored in the models (Figure~\ref{fig:negmodels}). A comparison of the observed distributions (Figures~\ref{fig:srti} and~\ref{fig:bati}) reveals that [Sr/Fe] values are significantly more similar to the negative $\kappa_{\rm X}$ for $M_{ESG}=10^{2} M_{\odot}$ models than the [Ba/Fe] values. However, it should be noted that we rule out the existence of a negative $\kappa_{\rm Ba}$ based on the MW halo data as the current UFD data are inconclusive on their own. In the next section, our results for ``allowed'' MDY strengths are compared with the most recent yields found the literature. 

\begin{table*}[t]
	\caption{Strength of Mass-Dependent Yields}
	\centering
ÊÊÊÊ\begin{tabular}{ccllll}
	\hline\hlineÊ\\[-2ex]ÊÊÊÊÊÊÊ
Element & Metallicity & $\kappa_{\it empirical }^{8-10M_{\odot}}$ ({\it r})$^{a}$& $\kappa_{\textit{ab initio}}^{15-40M_{\odot}}$  ({\it s})$^{b}$ & $\kappa_{\it inferred}^{15-40M_{\odot}}$ ({\it s})$^{c}$  &  {This Work} \\
            
(neutron-capture) & (Log Z) &  & (nr/rs) &  (rs/ss) &  \\ [0.5ex]
			\hline 
			\\ [-2.0ex]
ÊÊÊÊÊ&ÊÊÊ-5ÊÊÊÊÊÊÊÊÊÊÊÊ&  & $\sim$ 3.3/5.8 &Ê$\sim$ 6.5/6.7 & \\ [-1.5ex]
 \raisebox{2.0ex}{Strontium (Sr)}	&  Ê-3ÊÊÊÊÊÊÊÊÊÊÊÊ& \raisebox{2.0ex}{$\sim$ -15 or -18}Ê& $\sim$ 4.5/6.6 &Ê$\sim$ 7.4/---Ê&  \raisebox{2.0ex}{$(\lesssim -10),(\gtrsim 7)$}Ê\\ [0.5ex]

			\multicolumn{6}{c}{} \\
ÊÊÊÊÊÊ&ÊÊ-5ÊÊÊÊÊÊÊÊÊÊÊÊ&  &ÊÊÊ---ÊÊÊÊ&Ê$\sim$ 3.6/3.6  & \\ [-1.5ex]
 \raisebox{2.0ex}{Barium (Ba)} & Ê-3ÊÊÊÊÊÊÊÊÊÊÊÊ& \raisebox{2.0ex}{$\sim$ -15}  & ÊÊ---Ê &Ê$\sim$ 3.9/---Ê &  \raisebox{2.0ex}{$\sim(6-12)$}Ê\\ [1.0ex]
		\hline
		\multicolumn{6}{l}{$^a$ Derived from empirical yields given in \cite{cescutti12}.} \\
		\multicolumn{6}{l}{$^b$ Derived from Figure 4.14 of Frischknecht (2012, PhD Thesis) for non-rotating (nr)/rotating stars (rs). }\\
		\multicolumn{6}{l}{Yields for Ba were not given.} \\
		\multicolumn{6}{l}{$^c$ Derived from \cite{cescutti13} for rotating stars (rs) [their {\bf as}-models]/{\it spinstars} (ss) [their}\\
		\multicolumn{6}{l}{{\bf fs}-models].}\\
		\multicolumn{6}{l}{$^{\dag}$ \cite{chieffi04} and \cite{limongi12} provide another set of theoretical MDYs for Sr.}\\
		\multicolumn{6}{l}{From \cite{chieffi04} we find that the estimated MDYs for Sr given for progenitors with $z > 0$}\\
		\multicolumn{6}{l}{to $z\simeq z_{\odot}$ results in strengths that are $1\lesssim \kappa_{\rm Sr} \lesssim 4$. The MDY for Sr for zero metallicity stars is $\kappa_{\rm Sr} \simeq 8$}\\
		\multicolumn{6}{l}{ --- compatible with our work. However, more recent work by the same authors \citep{limongi12}}\\
		\multicolumn{6}{l}{produces a $\kappa_{\rm Sr}\lesssim 5$ for zero metallicity stars. This result is only marginally compatible with our findings.}\\
ÊÊÊÊ\end{tabular}
	\label{tab:mdy}
\end{table*}

\section{Discussion}\label{diss}
In this section, we evaluate how our model-derived MDY strengths compare to others found in the literature. We also examine how our selection of data affects our reported results.  

\subsection{Comparison to Other MDY Estimates}\label{comdy}
In Table~\ref{tab:mdy} we compare our derived MDY strengths to the latest predictions given in the literature. In particular, we compare our values to those extracted from {\it ab initio} yields (i.e. yields derived from simulations) for Sr given in Frischknecht (2012, PhD Thesis) and from inferred values from the {\it ab initio}- and {\it empirically}-derived yields (i.e. chosen to match observations) for Sr and Ba applied in \cite{cescutti13}.  

\begin{itemize}
\item {\bf \textit{Empirical} Yields for Sr and Ba ($8-10M_{\odot}$ production site)} In Cescutti \& Chiappini's work, their homogenous stochastic models are chosen to fit the general distribution of halo stars without binary enrichment. These models, which they refer to as {\it empirical} models, are employed by the authors to examine the distributions produced by applying both their {\it empirically}-determined MDYs for the {\it standard r}- (and {\it extended r}-)process sites and the newly derived {\it ab initio} yields from Frischknecht's thesis work. To generate MDY strengths for their {\it empirical} yields, we consult the figure of Sr and Ba yields given in \cite{cescutti12} which are reported to be similar to the yields used in \cite{cescutti13}.

\item {\bf \textit{ab initio} Yields for Sr ($15-40M_{\odot}$ production site)} In Frischknecht's work, he conducts a suite of simulations that produce various chemical yields from massive stars as a function of the stars' metallicity and rotation. From his work, we approximate {\it ab initio} strengths ($\kappa_{\textit{ab initio}}$) for $^{88}$Sr\footnote{In Frischknecht (2012, PhD Thesis), MDYs for Sr isotopes are said to show similar trends.} by examining Figure 4.14 of Frischknecht (2012, PhD Thesis). Unfortunately, we are unable to make a direct comparison to MDYs strengths for Ba (which are also evaluated by Frischknecht) because they are not available in his published work. 

\item {\bf \textit{Inferred} Yields for Sr and Ba ($15-40M_{\odot}$ production site)} We also generate an estimate of the MDYs for Sr, and more importantly, for Ba (unreported) from Frischknecht's unpublished results. To do this, we input the various inferred {\small $\Delta[\frac{X}{Fe}]$}, displayed in Figure 1 of \cite{cescutti13}, along with their progenitor stellar mass range into the difference between logarithmic values of Eqn.~\ref{eqn:mdy}. If we assume that Fe-yields for these stars are weakly mass dependent, we get:
\begin{equation}\label{eqn:delxfe}
\Delta\left[\frac{X}{Fe}\right] \sim Log \left(\frac{m_{X_{1}}}{m_{X_{2}}}\right) = \kappa_{X}\cdot Log \left(\frac{m_{1}}{m_{2}}\right)
\end{equation}
The estimates for the {\it inferred} MDYs strengths derived from Eqn.~\ref{eqn:delxfe} are also listed in Table~\ref{tab:mdy}.  
\end{itemize}

The final column of Table 1 gives our preferred MDY strengths, which are chosen by identifying ranges of $\kappa_{\rm X}$ that could be simultaneously compatible for BOTH the MW and UFD's (i.e. looking at both upper and lower panels) As seen in Figure~\ref{fig:results1a}, both positive and negative MDY strengths for Sr are allowed. In particular, both a $\kappa_{\rm Sr} \gtrsim 7$, consistent with Frischknecht's $15-40M_{\odot}$ {\it ab initio} yields and a $\kappa_{\rm Sr} \lesssim -14$, consistent with Cescutti's $8-10M_{\odot}$ {\it empirically}-derived ({\it standard  r}) yields, are favored for Sr. Additionally, the inferred $\kappa_{\rm Sr}$ from a combination of such yields {\it should}, in fact, be intrinsic to our analysis --- however, inferences about combined yields are beyond the scope of this investigation and shall be addressed in future work.

Figure~\ref{fig:results1b} shows us results for Ba yields. Positive MDYs with $\kappa_{\rm Ba} \sim 6-12$ are preferred and may be related to Frischknecht's {\it spinstar} yields. However, the extremely low likelihoods for negative $\kappa_{\rm Ba}$ when compared to  positive $\kappa_{\rm Ba}$, supports the notion that such yields are improbable. This strongly suggests a lack of Ba production from an $\sim8-10M_{\odot}$ production site which is consistent with more recent hydrodynamic simulations \citep[e.g.,][]{fischer10,wanajo11} but contrary to other expectations for nc-yields found in the literature \cite[see, e.g.,][]{cescutti13,cescutti12,qian08,wanajo03,ishimaru99,wheeler98,mathews92}.

These preliminary results illustrate the advantage of using statistical techniques that address the full density of the observed distributions and not only the average of their spreads as implemented in \cite{cescutti13} and other previous studies. Further development of this technique may provide the best chance to uncover the ``galactic genealogy'' of the MW and its closest companions in the Local Group.
\subsection{Effects of Data Selection on Results}\label{edsr}

\subsubsection{Data Compilations}\label{datcomp}
One concern about using a compilation of data such as \cite{frebel10} is that the dispersion in abundances may be artificially inflated by differences between the data sets. Frebel states that systematic difference between data sets are likely to inflate the dispersion by no more than 0.3 dex (for both the UFDs and MW halo).  In particular, the dispersion between different measurements of Ti abundance (i.e., via Ti {\scshape i}/Ti {\scshape ii} or a combination thereof) is typically, $\sim$0.1--0.15 dex \citep[e.g.,][]{shetrone03,aoki07,frebel10b} which is precisely on par with observational errors. 
In contrast, both the offset (under-abundance) and scatter (dispersion) of nc-elements are a factor of $\sim$3--5  and $\sim$10 bigger than these systematic uncertainties, respectively. Thus, we conclude that the differences between surveys cannot significantly alter our current results. Moreover, artificially inflated dispersions for UFD and MW halo distributions would serve to decrease their {\it expected} M$_{ESG}$ while leaving a significant $\Delta M_{ESG}$ between the distributions intact. Hence our result of lower M$_{ESG}$ for UFDs verses significantly higher M$_{ESG}$ for MW halo progenitors is insensitive to these systematic differences.

\subsubsection{Ignoring Data with Upper Limits}\label{idul}
Our parental likelihood test is not strictly applicable to samples containing upper limits. However, as a check, we apply the test to the Frebel data compilation, including upper limits, to determine the possible effects, if any, of leaving data with upper limits out of our analysis. Including the upper limits also increases the scatter of our MW halo samples, which, again, effectively decreases the inferred M$_{ESG}$ slightly while, in this case, increasing the inferred MDYs. These values are not significantly different from the values we report. The similarity of the results from the two samples is compatible with the fact all stars with only upper limits for Sr and Ba are consistent with having [Sr/Fe] and [Ba/Fe] abundance ratios above those stars with the lowest known levels of nc-elements \cite{roederer13}; which is to say that stars with upper limits would actually be detected if higher signal-to-noise spectra were available. Hence, star with upper limits are consistent with residing in, not below, the distributions of detected stars. The insensitivity (or compatibility) of the models to the exclusion or inclusion of data with upper limits proves that our work is sufficient for the purposes of broadly testing whether our simple scenario for chemical enrichment of UFDs in comparison to MW progenitors is plausible. Once the observed data sets for UFDs are larger a more rigorous statistical approach will be required to actually place strong limits on --- for example --- the detailed nature of MDY for nc-elements.

\section{Conclusion}\label{conc}
While the distribution of [Ti/Fe] is similar in both the MW halo and UFDs, the means/medians of nc abundance ratios for VMP stars found in these two systems are significantly offset. Although the current UFD sample is still small, this discrepancy motivates questions concerning the nature of hierarchical merging in the construction of the MW halo. In particular, discrepant abundance ratios suggest that past accreted dwarfs galaxies (i.e. progenitors of the stellar halo) may have been quite unlike the progenitors of the current MW satellites. Possible solutions include appealing to inhomogeneous chemical mixing, differential blowout of metals from SN winds, differences in primordial abundance ratios due to population III stars or differences in the IMF of stars within the progenitor systems.

In this paper, we explore an entirely different possibility for these discrepant abundance ratios: that progenitors of MW halo were enriched by a larger prior generation of stars when compared to UFD progenitors (as could be the case if, for example, UFD progenitors were more isolated than the MW progenitors, as suggested in \cite{corlies13}.
We demonstrate that this simple hypothesis can qualitatively and quantitively explain both the similarities of Ti distributions and differences between the nc-distributions for the current observed samples provided that the nc-elements have much stronger MDYs (currently unknown) than the (known) MDYs for Ti. Specifically, a viable model that simultaneously fits the distributions of [Ti/Fe], [Sr/Fe] and [Ba/Fe] is one in which MW progenitors were enriched by prior stellar generations of mass $M_{\rm ESG}\gtrsim 10^3 M_\odot$ and UFD progenitors were enriched by $M_{\rm ESG} \lesssim 10^2 M_\odot$. The most likely MDY strengths (given the data used and the simplicity of our models) are characterized by a power law index of $|\kappa_{\rm Sr}| \sim 7-14$ for Sr and $\kappa_{\rm Ba} \sim 6-12$ for Ba with lowest plausible values of $|\kappa_{\rm Sr,Ba}| \gtrsim 4$  (compared to $\kappa_{\rm Ti} \sim 1$). These numbers were derived from enriching stars sampled from for a Salpeter IMF with an upper limit of 40 $M_\odot$ (We show in Appendix~\ref{A1} that a different $m_{\rm upp}$ leads to a similar explanation, though with different numbers for $M_{\rm ESG}$ and $\kappa_{\rm X}$).\footnote{We anticipate that assuming a different form for the IMF (e.g. Kroupa IMF) would lead to a similar explanation with different but consistent parameter values.}

In this study we have demonstrated that our simple approach can explain the current data. However, it is known that many other effects can influence abundance ratio distributions in these systems, and, that ultimately, the relative importance of each effect needs to be assessed by building a more complete model. We see the current work as a foundation for more complete models in the future.

\begin{figure}[htbp]
   \centering
   \includegraphics[width=0.475\textwidth,angle=0]{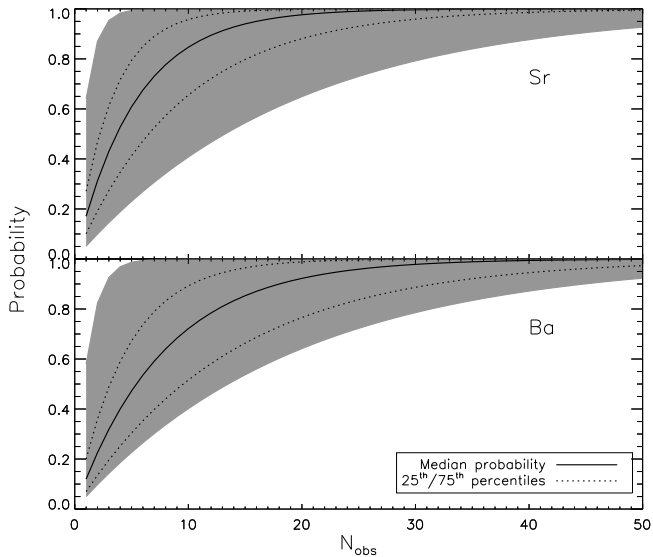}
   \caption{Gray region indicates the range in the probability that an observed UFD sample of size N$_{obs}$ could contain one star with [Sr/Fe] $> 0$ (top panel) or [Ba/Fe] $> 0$ (bottom panel) for parameter sets that had $p$-values greater than 0.05 (see lower panels of Figures \ref{fig:results1a} and \ref{fig:results1b}). The solid and dotted lines indicated the median and $25^{th}$/$75^{th}$-percentiles for these parameter sets, respectively.}
   \label{fig:results3}
\end{figure}

Despite the simplicity of our models, there are a number of interesting implications from our results. First, a relatively modest increase in the number of high-resolution spectra in UFDs could be used to test the specifics of our model --- if our interpretation is correct (barring other effects), then we should find UFD members with abundance ratios skewed above the bulk of the MW distribution as well as below. Figure~\ref{fig:results3} illustrates the likelihood of finding at least one UFD star with positive values of either [Sr/Fe] (upper panel) or [Ba/Fe] (lower panel) for different sample sizes. The gray region indicates the full range of probabilities for all parameter sets for which we found {\it p}-values $\ge 0.05$ when compared to the current data sets in Sections~\ref{rTiSr} and~\ref{rTiBa}. These probabilities were calculated from the parent distribution for each qualifying parameter set by finding the fraction of realizations, $f$, that had positive abundance ratios, and then adopting $f$ in the binomial theorem to estimate the probability of drawing at least one such star for sample size N$_{obs}$. The solid and dotted line indicates the median and $25^{th}$/$75^{th}$-percentiles for all qualifying parameters sets at a given N$_{obs}$. Overall, the figure indicates that, if our hypothesis of nc-abundance ratio distributions being skewed by strongly mass-dependent, power-law-like yields is a predominant effect, then sample sizes of $\sim$ 15 -- 25 VMP stars\footnote{If given a 95\% chance that a model produces at least one ``super-nc abundant'' star.} in UFDs should start to contain some nc-rich ([Ba,Sr/Fe] $\gtrsim 0$) counterparts to the nc-poor ([Ba,Sr/Fe] $< 0$) populations observed so far. Efforts made to extend stellar abundance ratio samples into the main sequence of UFDs \citep[e.g.][]{vargas13} should eventually provided samples large enough to determine whether stochastic sampling plays a predominant role in observed abundance ratio distributions. 

Second, it should also be noted that as sample sizes increase, the likelihood distributions in our parameter space will become more concentrated, providing stronger constraints on the form of MDY for nc-elements, and, by extension, their origin. A preliminary comparison of our current results with predictions for MDY in literature already suggests that while production of Sr from $8-10M_{\odot}$ stars is quite possible, production of Ba from these stars is highly unlikely. Our results also support the viability of recent {\it ab initio} yields for $15-40M_{\odot}$ stars.

In conclusion, our results indicate that abundance ratio distributions in nearby systems contain intriguing signatures of their early isolation (or conversely, contamination): more/less isolated systems should be enriched by smaller/larger prior enriching generations (i.e. to have lower/higher $M_{\rm ESG}$). These signatures could potentially be exploited to probe the progress of metal enrichment on MW scales in the early Universe --- a local window on a regime that cannot be seen directly.

\acknowledgements
KVJ thanks the Observatories of the Carnegie Institution of Washington for their hospitality during a visit which provided the inspiration for this work, and Andrew McWilliam and Ian Thompson for conversations on this topic in particular. DL would also like to acknowledge the encouraging and insightful conversations he had with Ian Roederer, Andrew McWilliam, Alan Dressler, Chris Sneden, George Preston, Luis Vargas, Tim Beers, Anna Frebel, and Volker Bromm. Finally, we would like to thank the referee for comments which helped to broaden the scope of our work to compare our mass-dependent yields with those found in recent literature. DL was supported by NSF grants AST-0806558 and AST-1107373.

\bibliography{reading1,allrefs}

\appendix
\section{{\bf A \hspace{5pt}  Effects of Varying the Upper Mass Limit}} \label{A1}
\begin{figure}[htbp]
   \centering
   \includegraphics[width=.95\textwidth,angle=0]{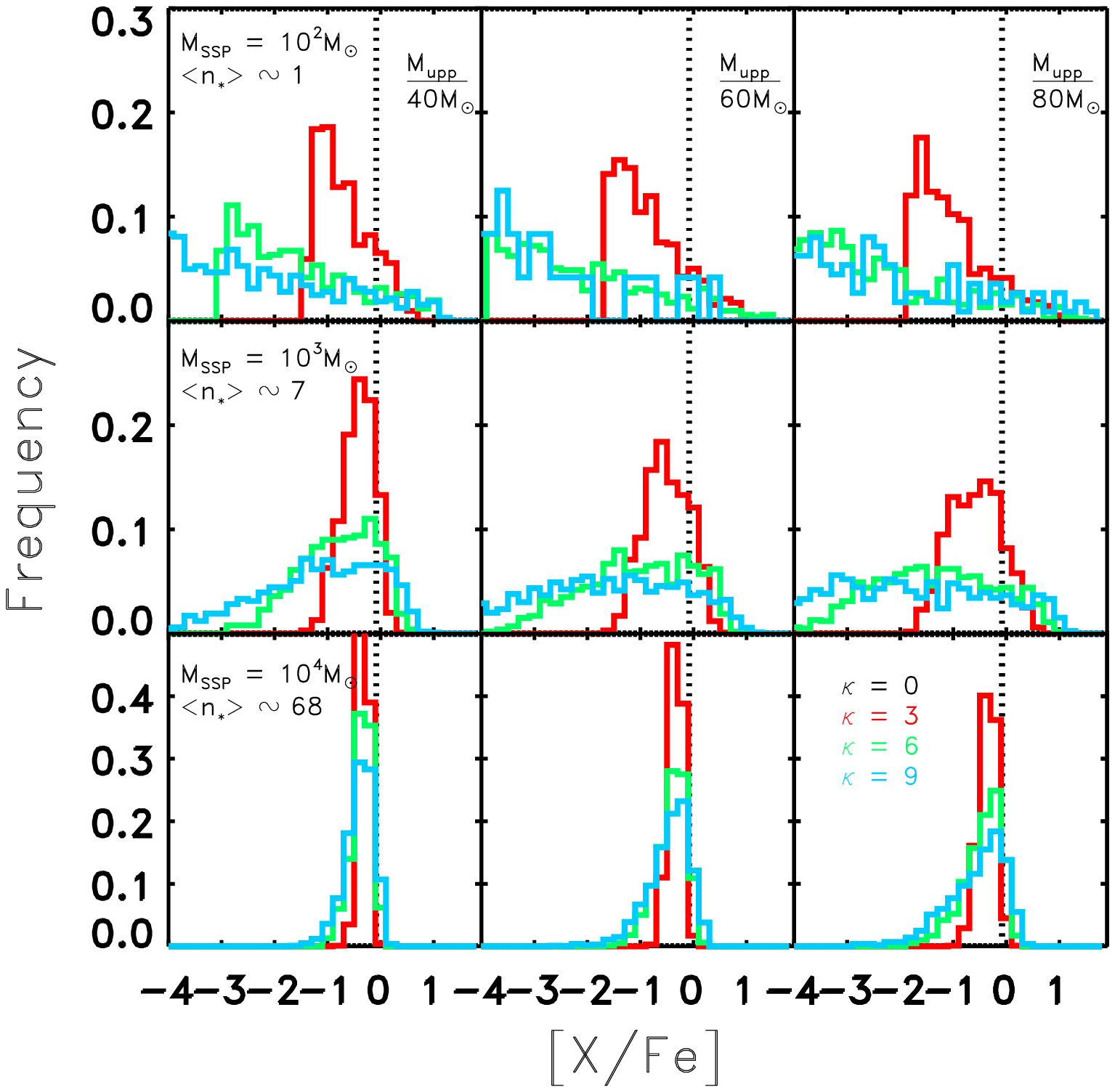}
   \caption{Distributions of abundance ratios produced from 1000 realizations of an ESG, with $M_{\rm ESG} = 10^{2} M_{\odot}$ (top row), $10^{3} M_{\odot}$ (middle row), and $10^{4} M_{\odot}$ (bottom row). Each column represents models generated with different $m_{\rm upp}$ for the IMF: $40 M_\odot$ (first column), $60 M_\odot$ (second column), and $80 M_\odot$ (third column). Colors are the same as found in Figure~\ref{fig:models} of the paper.}
   \label{fig:models2}
\end{figure}

\begin{figure}[htbp]
   \centering
   \includegraphics[width=.95\textwidth,angle=0]{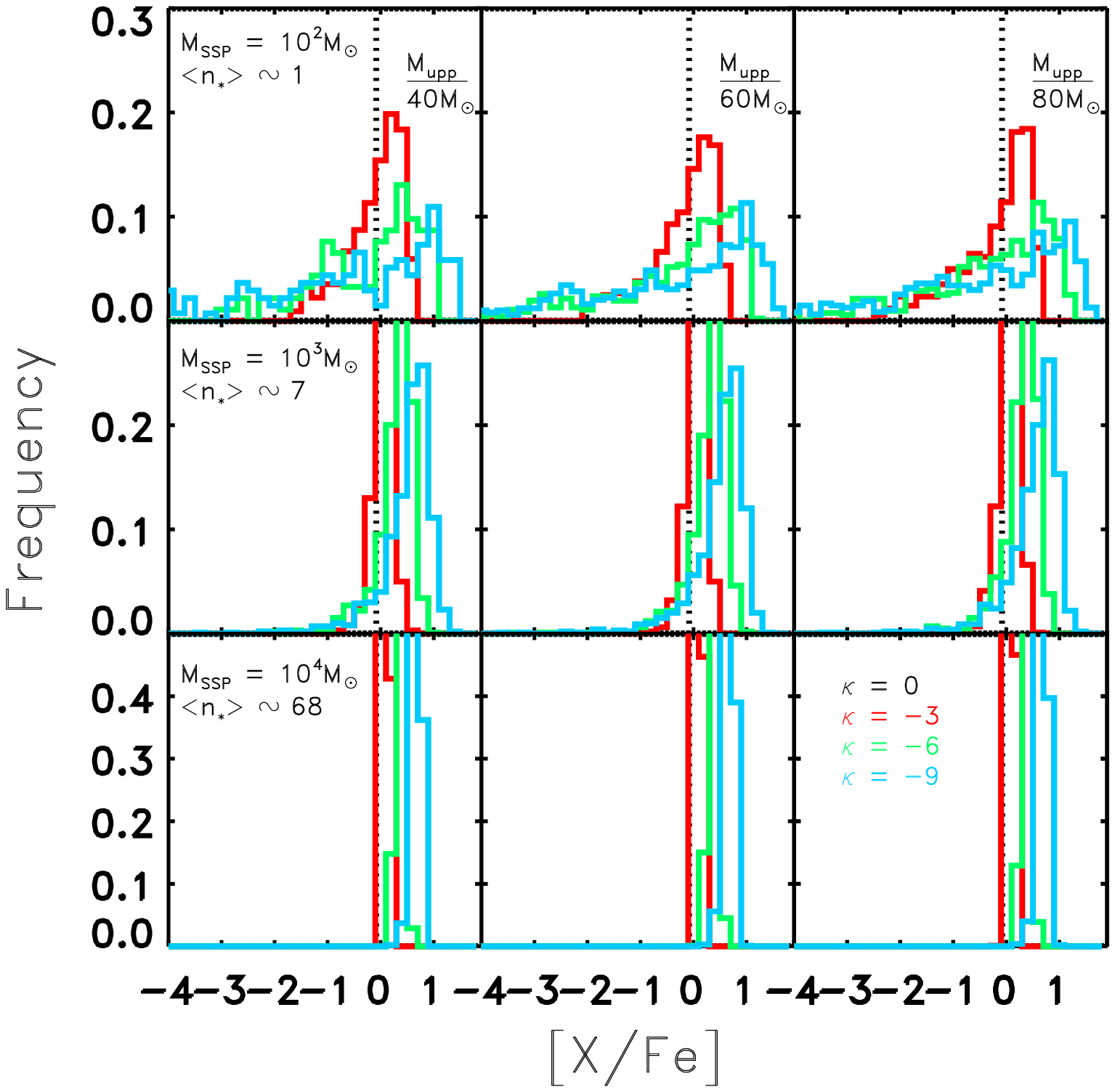}
   \caption{Figure is the similar to Figure~\ref{fig:models2} but shows distributions derived from negative MDYs.}
   \label{fig:negmodels2}
\end{figure}

 The plots in this appendix illustrate the results of adopting a different assumption for the high-mass cutoff of the IMF. Figure~\ref{fig:models2} repeats Figure~\ref{fig:models} (left panels), adding panels for $m_{\rm upp} = 60$ (middle panels) and 80 (right panels) to demonstrate how the yield distributions are affected across various $m_{\rm upp}$. As $m_{\rm upp}$ increases, the range in possible yields is increased, stretching the distribution in a similar manner to increasing $\kappa_{\rm X}$. While there is some noticeable degeneracy in the effects of $m_{\rm upp}$ and $\kappa_{\rm X}$ on skewness and in the effects of $m_{\rm upp}$ and $M_{\rm ESG}$ on dispersion, simple inspection suggests that the change in $m_{\rm upp}$ most strongly affects the kurtosis (peakedness) of the distribution. Increasing $m_{\rm upp}$ widens and flattens the distributions seen across the columns of Figure~\ref{fig:models2}. Figure~\ref{fig:negmodels2} displays a similar profile for negative MDYs.

\begin{figure}[htbp]
   \centering
   \includegraphics[width=.95\textwidth,angle=0]{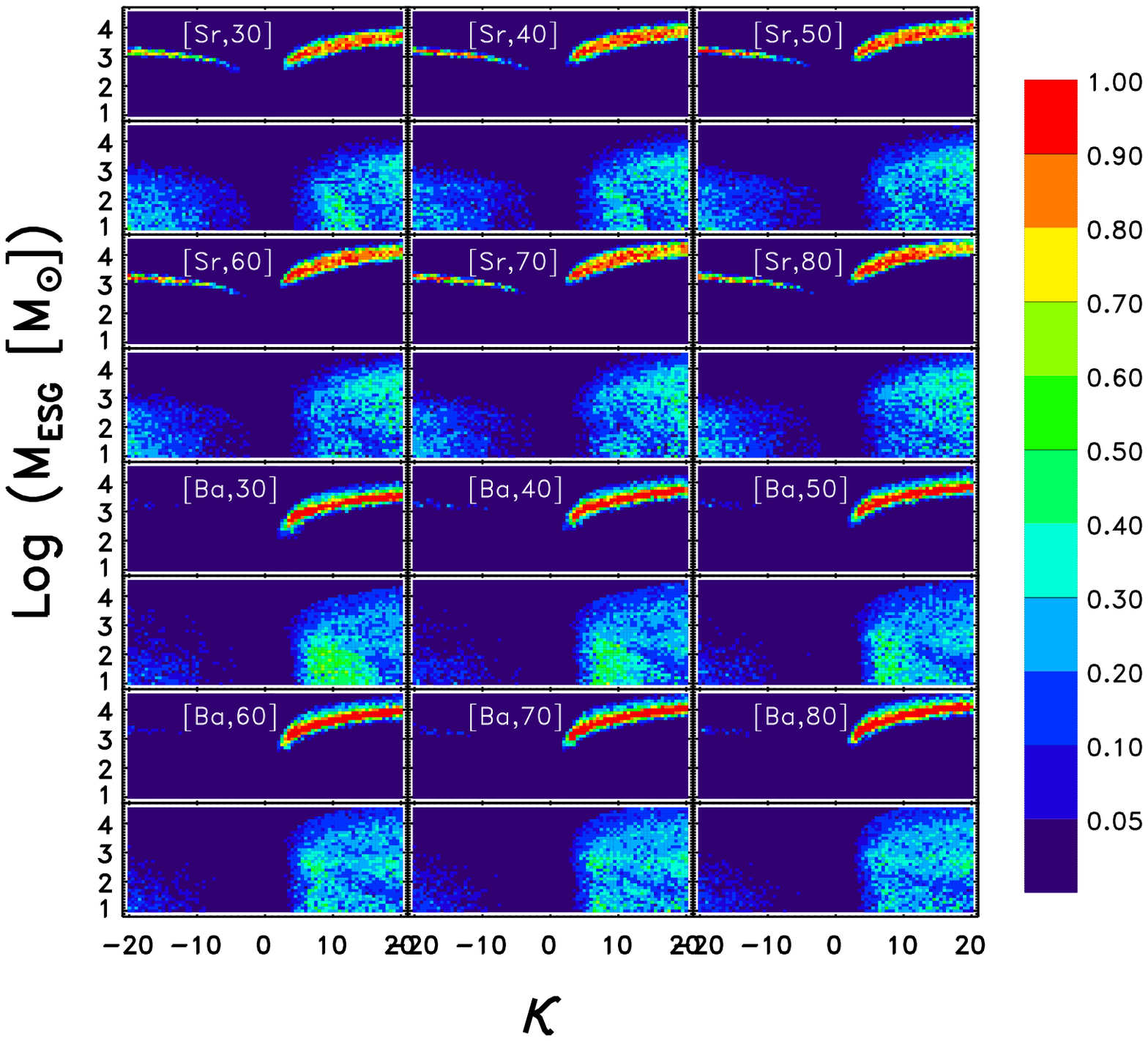}
   \caption{Likelihoods ({\it p}-values) for all models examined as a function of $M_{\rm ESG}$, $\kappa_{\rm X}$, and $m_{\rm upp}$ for the two abundance ratio planes investigated. Models are indicated by a ``Sr'' for the [Ti/Fe]-[Sr/Fe] plane or a ``Ba'' for  [Ti/Fe]-[Ba/Fe] plane and a value for $m_{\rm upp}$ in solar masses.}
   \label{fig:results2}
\end{figure}

Figure~\ref{fig:results2} displays $p$-values from our paternal likelihood test for a variety of $m_{\rm upp}$ in both the [Ti/Fe]-[Sr/Fe] and [Ti/Fe]-[Ba/Fe] planes. As $m_{\rm upp}$ increases, our probability distributions shift to higher $M_{\rm ESG}$ values for a given $\kappa_{\rm X}$: the wider range of stellar masses means a wider range in individual SNe yields for a given $\kappa_{\rm X}$, requiring a larger $M_{\rm ESG}$ to match the observed abundance ratio spreads. However, the figure confirms that our general results - of the UFD distributions requiring large $|\kappa_{\rm X}|$ and smaller $M_{\rm ESG}$ than the MW distributions - are robust despite our ignorance of the actual value of $m_{\rm upp}$. 

\section{{\bf B \hspace{5pt}  Derivation of our MDY models}} \label{A2}
To create an analytic average abundance ratio from Eqn.~\eqref{eqn:mdy}, we first calculate a mass ratio in the limit of a completely sampled IMF:
\begin{align}
\label{eqn:mxfe}
\frac{M_{X}}{M_{Fe}} & = \frac{\int_{m_{\rm enrich,low}}^{m_{\rm upp}} \! \xi\cdot m_X \, dm}{\int_{m_{\rm enrich,low}}^{m_{\rm upp}} \! \xi\cdot m_{Fe} \, dm} =   \left(\frac{\beta_{\rm X}}{\beta_{\rm Fe}}\right)\cdot\left(\frac{\kappa_{\rm Fe} - \alpha +1}{\kappa_{\rm X} - \alpha +1}\right)\cdot\left(\frac{ {m_{\rm upp}}^{\kappa_{\rm X} - \alpha +1} - {m_{\rm enrich,low}}^{\kappa_{\rm X} - \alpha +1} }{ {m_{\rm upp}}^{\kappa_{\rm Fe} - \alpha +1} - {m_{\rm enrich,low}}^{\kappa_{\rm Fe} - \alpha +1} } \right) 
\end{align}
where M$_{X}$ and M$_{Fe}$ represent the total mass yield in X and Fe. 

We can relate Eqn.~\eqref{eqn:mxfe} to solar abundance ratios and thereby calculate the IMF-weighted chemical abundance ratio [X/Fe]:
\begin{align}
\label{eqn:avgxfe}
\left[\frac{\rm X}{\rm Fe}\right]_{\rm IMF} \equiv & \log \left(\frac{M_{X}}{M_{Fe}}\right) - \log \left(\frac{M_{mol}(X)}{M_{mol}(Fe)}\right) - \log \left(\frac{N_{X,\odot}}{N_{Fe,\odot}}\right) 
\end{align}
where {\small $\left[\frac{\rm X}{\rm Fe}\right]_{\rm IMF}$} is the IMF-weighted abundance ratio, {\small $\frac{M_{mol}(X)}{M_{mol}(Fe)}$} is a ratio of the molar masses of X and Fe, and {\small $\frac{N_{X,\odot}}{N_{Fe,\odot}}$} is the ratio of the solar abundances of X and Fe. We can then use the average of observed abundance ratios from our sample of MW halo stars, {\small $\left<\left[\frac{\rm X}{\rm Fe}\right]\right>_{\rm OBS}$}, to calibrate $\beta$ for a given $\kappa_{\rm X}$ of element X by using Eqn.~\ref{eqn:xfe}.


\end{document}